# Relativistic Tennis with Photons: Demonstration of Frequency Upshifting by a Relativistic Flying Mirror through Two Colliding Laser Pulses[*]

M. Kando[1], Y. Fukuda[1], A. S. Pirozhkov[1,2], J. Ma[1], I. Daito[1], L.-M. Chen[1], T. Zh. Esirkepov[1,3], K. Ogura[1], T. Homma[1], Y. Hayashi[1], H. Kotaki[1], A. Sagisaka[1], M. Mori[1], J. K. Koga[1], H. Daido[1], S. V. Bulanov[1,3,4], T. Kimura[1], Y. Kato[1] and T. Tajima[1]

[1]*Advanced Photon Research Center, Japan Atomic Energy Agency, Kizugawa, Kyoto, Japan*

[2]*P. N. Lebedev Physical Institute, Russian Academy of Sciences, Moscow, Russia*

[3]*Moscow Institute of Physics and Technology, Dolgoprudny, Moscow region, Russia*

[4]*A. M. Prokhorov Institute of General Physic, Russian Academy of Sciences, Moscow, Russia*

**Abstract:** Since the advent of chirped pulse amplification[1] the peak power of lasers has grown dramatically and opened the new branch of high field science, delivering the focused irradiance, electric fields of which drive electrons into the relativistic regime. In a plasma wake wave[2] generated by such a laser, modulations of the electron density naturally and robustly take the shape of paraboloidal dense shells,[3,4] separated by evacuated regions, moving almost at the speed of light. When we inject another counter-propagating laser pulse, it is partially reflected from the shells, acting as relativistic flying (semi-transparent) mirrors, producing an extremely time-compressed frequency-multiplied pulse which may be focused tightly to the diffraction limit.[5] This is as if the counterstreaming laser pulse bounces off a relativistically swung tennis racket, turning the ball of the laser photons into another ball of coherent X-ray photons but with a form extremely relativistically compressed to attosecond and zeptosecond levels. Here we report the first demonstration of the frequency multiplication with a factor $\approx 50\ldots100$ detected from the reflection of a weak laser pulse (source pulse) in the region of the wake wave generated by the driver pulse in helium plasma. This leads to the possibility of very strong pulse compression and extreme coherent light intensification.[5,6] This "relativistic tennis" with photon beams is demonstrated leading to the possibility toward reaching enormous electromagnetic field intensification and finally approaching the Schwinger field, toward which the vacuum nonlinearly warps and eventually breaks, producing electron-positron pairs.

## 1. Introduction

The development of laser technology has resulted in a tremendous growth of the light intensity in the laser focal spot. Electrons in the laser electromagnetic field become relativistic at intensities I ≈ $2\times10^{18}$ W/cm$^2$. Nowadays lasers produce focused irradiance approaching $10^{22}$ W/cm$^2$ (Ref. 7), which makes plasma dynamics ultrarelativistic. By increasing irradiance further we shall encounter novel physical processes such as the Compton scattering dominated laser-matter interaction, and then, at irradiances of the order of $10^{29}$ W/cm$^2$, the focused light can generate electron-positron pairs in vacuum.[8,9] Several ways have been suggested to achieve such intensity (see articles, Refs. 5,6, 10-12, and literature quoted in). Here we consider the "flying mirror" concept:[5] at optimal conditions, the dense shells formed in the electron density in a strongly nonlinear plasma wake, generated by a short laser pulse, reflect a portion of a counter-propagating laser pulse. The reflected radiation is frequency-upshifted due to the double Doppler effect, as predicted by A. Einstein,[13] and focused due to the paraboloidal shape of the shells, caused by the relativistic dependence of the plasma frequency on the wake wave amplitude.[3,4] This leads to the pulse shortening and light intensification by a large amount. In this concept it is essential that the wake wave must be close to wave-breaking. If the wake wave is far below the wave-breaking threshold, the reflection of the counter-propagating laser pulse at the electron density modulations is exponentially small. Near this threshold, the reflection becomes much more efficient,[5,6] because the electron density of the shells takes the profile of cusps,[14] moving with phase velocity, $v_{ph}$, close to the speed of light $c$. At the wave-breaking point the Lorentz factor $\gamma_e$ of electrons forming the shells becomes equal to $\gamma_{ph}= (1 − \beta_{ph}^2)^{-1/2}$, where $\beta_{ph}= v_{ph}/c$. Since the wake wave is generated by the driver pulse, the factor $\gamma_{ph}$ is close to $\omega_0/\omega_p$ (Refs. 2,14), where $\omega_0$ is the laser frequency, $\omega_p=(4\pi n_e e^2/m_e)^{1/2}$ is the plasma frequency, $n_e$ is the electron density, and $e$ and $m_e$ are the electron charge and mass. The frequency of the reflected light is multiplied and the pulse duration is shortened by a factor $\omega_X/\omega_0 = (1+\beta_{ph}^2+2\beta_{ph}\cos\theta)\,\gamma_{ph}^2$, where $\theta$ is the incidence angle.

## 2. Laser system and diagnostics

The proof-of-principle experiment of the "flying mirror" concept is performed by colliding two 76 fs laser pulses in a helium supersonic gas jet. The experiment setup is shown in Fig.1.

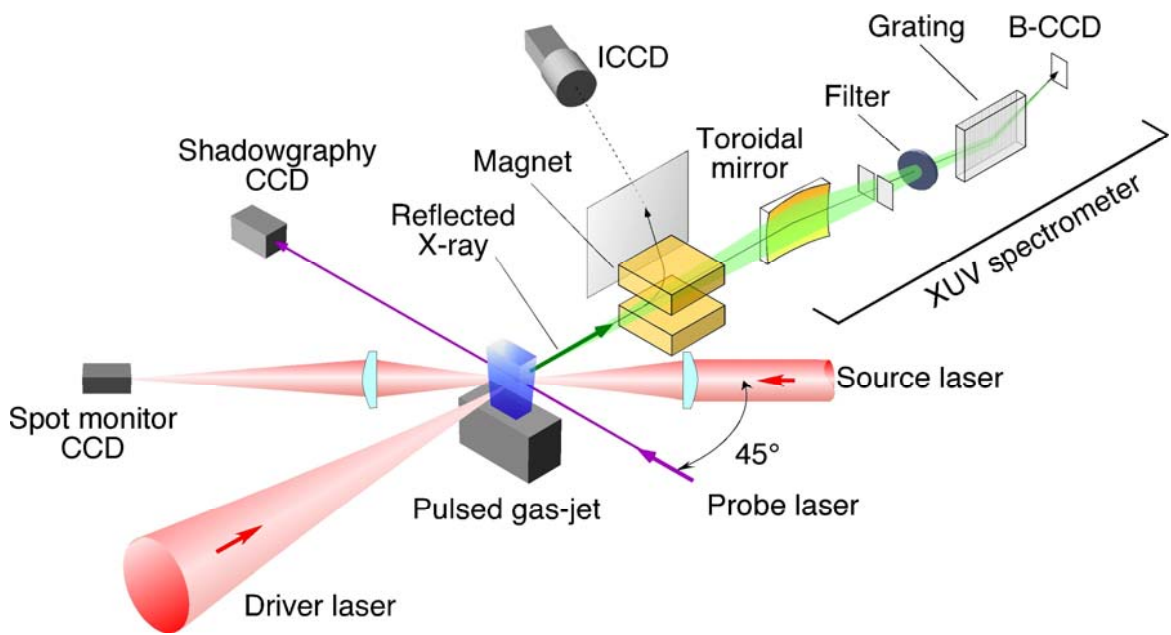

**Figure 1. Setup of the flying mirror experiment.** The driver laser is focused into the He gas jet to make a wake, which forms the "flying mirror" when it reaches the breakpoint. The source laser is focused to the wake breaking region at the incidence angle θ = 45°. The probe laser is used for precise alignment. The magnet bends accelerated electrons which hit the phosphor screen; the electron spectrum is monitored with the intensified CCD. The signal reflected by the wake is measured with a grazing incidence XUV spectrometer.

We used the JLITE-X laser at the Advanced Photon Research Centre, Japan Atomic Energy Agency. The laser produced 210 mJ, 76 fs pulses at the center wavelength of 820 nm. The horizontally polarized driver laser pulse was focused by an off-axis parabolic mirror with a focal length of 645 mm into a supersonic gas jet (the slit is 10 mm by 1.26 mm; the driver pulse propagated along the short side of the slit). The source pulse was focused by a plano-convex lens with a focal length of 200 mm, placed on a 5-axis movable stage in order to achieve spatial overlapping of the laser pulses. The driver pulse had a $1/e^2$ focal spot diameter of 27 μm and the estimated irradiance in vacuum of $5\times10^{17}$ W/cm$^2$. The source pulse energy was 6.3% of the driver pulse, and an estimated irradiance in vacuum of ~ $10^{17}$ W/cm$^2$. The fast electrons, generated in the laser–plasma interaction, bent by a permanent magnet, hit the phosphor screen, which was monitored with an intensified CCD. The spatial overlap of the two laser pulses was arranged using a reference needle, placed in the focus region, and two CCD monitors. The two laser pulse collision and channeling in the plasma were observed in the shadowgram and interferogram produced by the probe laser beam. The signal reflected by the wake wave was measured with a grazing incidence XUV spectrometer composed of a toroidal mirror, a diffraction grating, and a back-illuminated CCD. Two optical blocking filters (free-standing Mo/C multilayer stacks with 60 periods 2.6 nm thick each) were used to block the driver laser light. The spectrometer was calibrated using line emission from Ar and Ne gas targets in the wavelength range from 4.6 nm to 15.8 nm. The spectral resolving power was $\lambda/\Delta\lambda=100$ (200) at 5nm (15 nm). The spatial resolution in the vertical direction was approximately equal to 60 μm. The acceptance angle was $10^{-4}$ sr.

## 3. PIC Simulations

To understand the details of the interaction under the present experimental conditions, we performed two-dimensional Particle-in-Cell (PIC) simulations. The Particle-in-Cell (PIC) simulations were performed using the Relativistic Electro-Magnetic Particle-mesh code (REMP) running on the Altix 3900 supercomputer at JAEA-Tokai.

The simulations were carried out for the full-scale plasma and, with much higher resolution, for the smaller region close to the point of intersection of the two pulses. The former case appeared to be in good agreement with the patterns seen in shadowgraphy and electron acceleration observations and confirmed the choice of the intersection point (see discussion below). The plasma density profile and laser intensity were set close to the conditions of the experiment. We observed a strong modulation of the laser pulse after 1-1.5 ps of propagation towards the gas jet center and, as a result, an excitation of a wakefield with a regular structure, yet close to wave-breaking, in a region 200-300 μm before the gas jet and the formation of solitons near the jet center, in agreement with experimental findings. The collision of two laser pulses was then simulated in a smaller region of the plasma, represented by $1.5 \times 10^8$ quasiparticles, with much higher resolution (512 and 64 meshes for the laser wavelength in the longitudinal and transverse direction, respectively) to reveal high frequency upshifts.

The result of the latter case is shown in Fig. 2: the driver pulse excites a wakefield near the wave-breaking, the dense shells acting as "flying mirrors" partially reflect the incident source pulse, whose large frequency upshift and anisotropy upon reflection are prominently evident. The reflected emission is estimated as $5.2 \times 10^{10}$ photons/sr, in agreement with theoretical prediction.

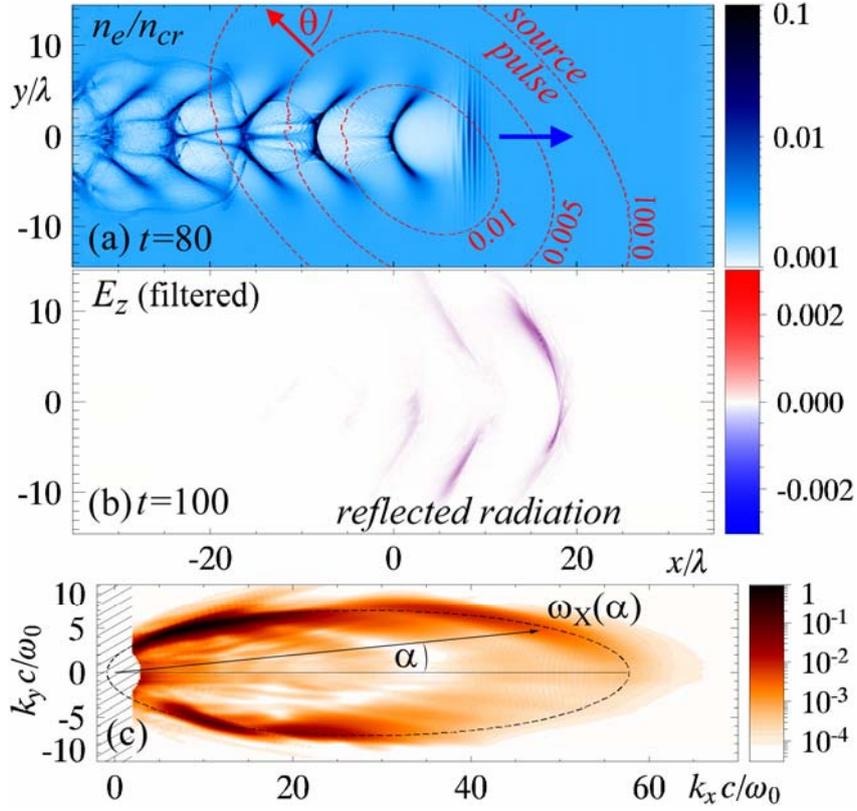

**Figure 2. The PIC simulation results.** (a) The normalized electron density in the wake of the driver laser pulse and the source pulse dimensionless amplitude contours (dashed curves). (b) The dimensionless electric field component $E_z$ of the radiation reflected from plasma dense shells (all frequencies $\leq \omega_0$ are filtered out). Time is in laser periods. (c) Spectral intensity of the reflected radiation (in arbitrary units); the hatched region is filtered out. The point $(k_x, k_y)$ corresponds to the frequency of radiation emitted at the angle $\alpha = \arctan(k_y/k_x)$, $\omega_X(\alpha) = c(k_x^2 + k_y^2)^{1/2}$. The dashed ellipse is for the theoretical dependence of the reflected frequency as function of $\alpha$: $\omega_X(\alpha) = \omega_0 (1 + \beta_{ph}\cos\theta)/(1 - \beta_{ph}\cos\alpha)$ at $\gamma_{ph} = 4.17$.

## 4. Results

A 2 TW driver laser pulse is used to excite the plasma wake wave, while the source pulse, obtained by splitting the main laser pulse, is focused onto the wake region. To ease the setup, the source pulse is incident at an angle of $\theta = 45^\circ$ with respect to the reflecting dense shells in the wake wave. In this configuration the theoretical frequency multiplication factor is $\omega_X/\omega_0 \approx 3.4\gamma_{ph}^2$. The plasma density is set to be about $5\times10^{19}$ cm$^{-3}$. The chosen parameters of the driver pulse and gas jet secure the necessary optimal wake wave excitation, since in previous experiments[15] quasi-monoenergetic electron beams were observed under this condition. We used the fast electron acceleration, of the order of 20 MeV in our case, as evidence of the breaking plasma wake wave formation. Although the mono-energetic electron beam generation is useful indication of the wake wave breaking, strictly speaking, it is not required for the relativistic mirror formation. Additional evidence of the plasma wave excitation came from the analysis of the scattered driver pulse spectrum observed at 60°, which exhibited red- and blue-shifted maxima, corresponding to Stimulated Raman Scattering. The interferogram showing the plasma channel is also consistent with the formation of the wake wave leading to the electron density redistribution.

In the experiment, it was crucially important to properly choose the colliding point position inside the plasma. The position of the two beams was determined by the shadowgram obtained with the third probe laser beam, as seen in Figs. 1 and 3a. Easing of the source laser pulse aiming onto the wake region was also achieved by observing the stationary, well localized structures inside the self-focusing channel near the centre of the gas jet (Fig. 3b). These structures were identified as post-solitons, the late stages of the evolution of relativistic electromagnetic solitons.[16-18] They appear where the laser pulse energy is substantially depleted and thus indicate the limit to which the good quality wake extends.

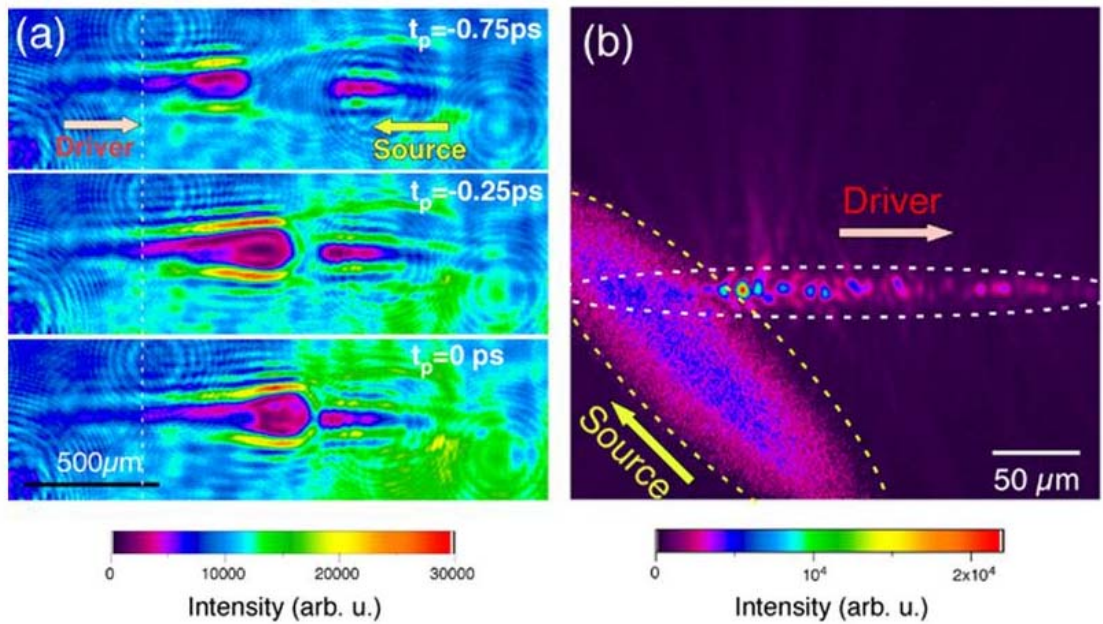

**Figure 3. Two colliding laser pulses seen in monitors.** (a) Side view (shadowgraph by the probe beam). The driver pulse propagates from the left to the right; the source pulse direction is opposite. The laser pulse collision occurs around the delay time $t_p$= 0 ps. (b) Top view image (obtained with a narrow-band 800 nm filter): the colliding laser pulses in their mutual plane. This is a superposition of two images obtained from shots with the source pulse alone and with the driver pulse alone. (Both could not be seen in one shot since the source

pulse was much weaker). The bright spots within the driver path correspond to post-solitons.

The temporal overlapping with a picosecond accuracy was controlled by the analysis of the transmitted source spectrum. The blue shift caused by ionization effects[19,20] was observed irrespective of the driver pulse presence. However, a delay-dependent additional blue shift was observed when the source pulse intersected the wake wave region, due to the fast change of the refractive index there.

Extreme accuracy (a few microns × tens femtoseconds) was required to point the source laser pulse at the location of the wake wave breaking, where the "flying mirror" is formed. This was met by controlling the targeting by observing a bright spot in the scattered light of the unfocused source pulse (the details of the technique will be published elsewhere). This 15-20 μm spot moving almost at the speed of light is interpreted as the electron density modulation, associated with the driver pulse and a wake region behind it, leading to the source pulse refraction. Aiming at this *moving spot*, we varied the delay time between pulses τ and the vertical position $z$ of the source pulse. In this way a wide range of colliding point coordinates was scanned. Using the XUV spectrometer, we detected 24 highly relevant signals of the reflected radiation in different shots (Fig. 4). The majority of signals was seen when the colliding point was located 200 μm before the gas jet centre (the driver focal plane was located 600 μm before the gas jet centre).

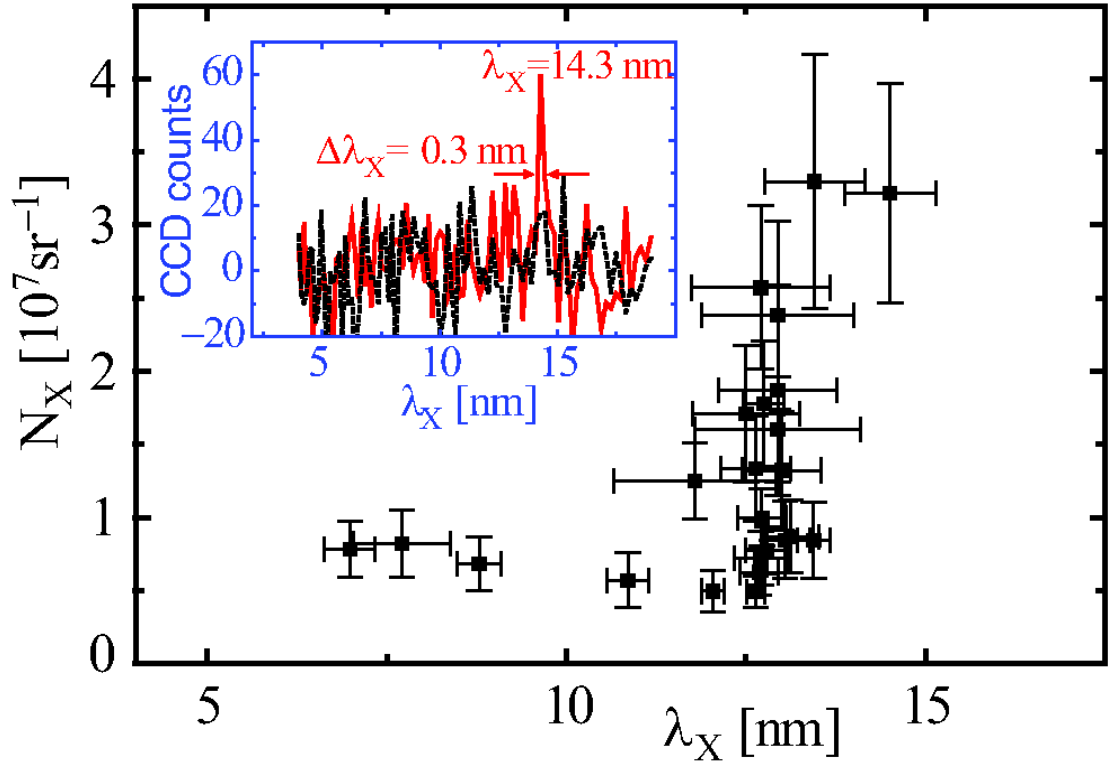

**Figure 4. Detected XUV signals in different shots.** Each point corresponds to radiation reflected in plasma in a single shot. The horizontal bars refer to the spectral width, the vertical bars correspond to the photon number uncertainty (see text). **Inset:** the XUV spectrometer readout for one of the signals (solid line) and the readout for the case without the source pulse (dashed line).

The inset in Fig. 4 shows the XUV spectrometer readout for the signal corresponding to the reflected radiation wavelength $\lambda_X = 14.3$ nm, which is 56 times shorter than the incident source pulse wavelength. The signal bandwidth is $\Delta\lambda_X = 0.3$ nm and the detected photon number per shot is $N_d = 25 \pm 7$. The uncertainty in the photon number arises from both the shot noise and the continuous background plasma emission. Using idealized values for the toroidal mirror reflectivity $R_T = 0.8$, the transmission of two filters $T_F = 0.3^2$, the diffraction grating efficiency $\eta_G = 0.2$ and the CCD quantum efficiency $\eta_{CCD} = 0.4$, we can estimate the number of photons reflected by the relativistic "flying mirror" formed in the plasma wake wave. This amounts to the emission of $3.3 \times 10^7$ photons/sr. The detected value is lower than the theoretically expected number and the reflected photon number seen in the PIC simulations because of the non optimal conditions of the collision (at the interaction point, the source pulse intensity is ~30 times lower due to propagation in the gas jet). Not taken into account here are effects of contamination, spectrometer optics roughness, etc., which can substantially increase the estimated number of reflected photons.

The detected wavelength and shortening factor $\lambda_0/\lambda_X \approx 3.4\gamma_{ph}^2$ (due to reflection from the "flying mirror") give the Lorentz factor $\gamma_{ph}^{FM} = 4.1$. We can also estimate this factor from the observed fast electron bunch energy, $E_b = m_e c^2 \gamma_b \approx 2 m_e c^2 \gamma_{ph}^2$ (Ref. 2). From observed 19 MeV electrons in this shot, we obtain $\gamma_{ph}^b = 4.3$, in good agreement with previous estimation. Assuming that the reflected signal has inherited the coherence from the source pulse, we find its estimated duration to be 1.4 fs, which is consistent with the observed spectral width $\Delta\lambda_X/\lambda_X \approx 1/48$. Our simulations also match well with the spectral characteristics such as the value of $\gamma_{ph}$, $\lambda_X$, and $E_b$. These independent factors reinforce each other.

The detected photon number correlates with the accuracy of targeting, as shown in Fig. 5. This correlation imposes constraints on the vertical size of the "flying mirror" and the duration of the incident radiation (before reflection), which are consistent with the expected transverse size of the wake wave and the source pulse duration.

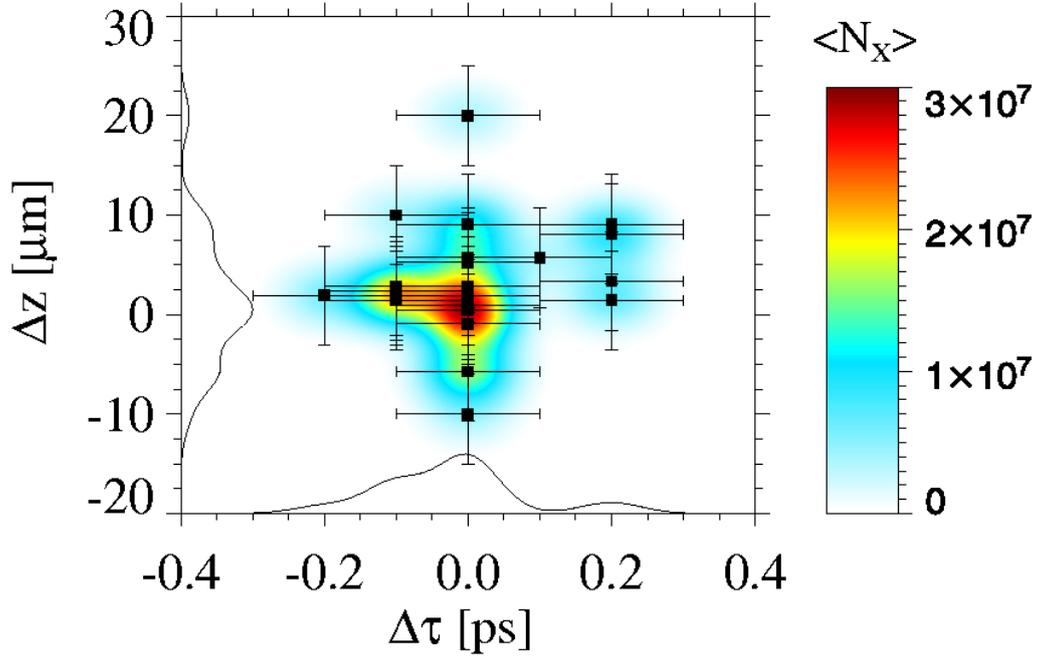

**Figure 5. The distribution of the photon number density** in the plane ($\Delta\tau$, $\Delta z$), where $\Delta\tau$ and $\Delta z$ determine deviations from the colliding point, predicted using the *"moving spot"* observation (see text). Unit of <Nx> is 1/sr/ps/μm. Error bars are determined by the accuracy of the monitors.

Without the source pulse, we observed continuous bremsstrahlung plasma emission with a vertical size equal to 200…300 μm. In contrast, the signal reflected by the "flying mirror" was highly localized in space, which distinguished it from the continuous emission of plasma. The measured vertical signal size was 60 μm, which was essentially limited by the spatial resolution of the XUV spectrometer; we believe, the real size was much smaller, of the order of the self-focusing channel width ($\leq 10$ μm), as can be seen from Fig. 5. The contribution from the high-order harmonic generation is negligible because helium becomes fully ionized well before the driver pulse peak arrival (the barrier-suppression-ionization threshold for He$^{2+}$ is about $3.3\times10^{16}$ W/cm$^2$). The signal spectrum (Fig. 4, inset) is shown after the background has been subtracted. The background is a 20-shot average of spectra obtained without the source laser pulse. The measured noise (standard deviation) for a single pixel when the source pulse was blocked is $\sigma_1$ = 2.38±0.04 counts; 0.04 is the standard deviation of the shot-to-shot $\sigma_1$ fluctuations. This is in good agreement with the calculated value, determined by the CCD design, $\sigma_{1C}$ = 2.40±0.03, obtained from the measured readout noise $\sigma_r$ = 1.897±0.014 and the shot noise $\sigma_s$ = $(gC)^{0.5}$ using the relation[22] $\sigma_{1C}^2$ = $(1+1/20)(\sigma_r^2 + \sigma_s^2)$. Here $g$ = 0.315 counts/electron is the CCD gain, $C$ = 5.9±0.4 is the observed background counts per pixel, the coefficient (1+1/20) is due to the subtraction of the 20-shot averaged background. For the spectrum binned by 12×2 pixels (Fig. 4, inset), the noise is $\sigma$ = $(24)^{0.5}\sigma_1$ = 11.7±0.2 counts. The observed signal peak (60 counts) is five times larger than the standard deviation of background fluctuations.

We note that the small bandwidth of the detected signals hinders explaining the reflection in terms of Thomson scattering. The contribution from the Thomson scattering off fast electrons is orders of magnitude smaller than the detected number of photons within the given spectral range and its spectrum is much broader.[21] For the detected Lorentz factor the reflection on the bulk electrons occurs mainly at dense shells in the electron density in a collective manner, as described in the "flying mirror" concept.[5]

## 5. Conclusions

When we collate all the measurements and supporting simulations, we realize that the evidence from each measurement reinforces each other, thus leading to the rigid conclusion of the first ever observation of a collective relativistic frequency multiplier through two colliding laser pulses in plasma. In these experiments the proof-of-principle of the "flying mirror" concept has been demonstrated. In the future this effect will allow the production of tunable sources of hard electromagnetic radiation with controlled parameters desired for a wide range of applications.[6,23]

This meets the *relativistic engineering* paradigm, in which we consciously exploit the consequences of relativity that matter moves more coherently and in a more controlled way because of its synchronicity with the speed of light in this regime.

**Acknowledgments:** This work was supported by the Ministry of Education, Science, Sports and Culture of Japan, Grant-in-Aid for Specially Promoted Research No. 15002013. We acknowledge useful discussions with I. A. Artiukov, M. Borghesi, M. Fujiwara, T. Hayakawa, T. Kawachi, V. I. Luchin, M. Koike, K. Krushelnick, A. Maksimchuk, V. Malka, K. Mima, A. V. Mitrofanov, G. A. Mourou, K. Nakajima, K. Nishihara, F. Pegoraro, E. N. Ragozin, N. N. Salashchenko, V. Yanovsky, M. Zepf, and A. G. Zhidkov. We thank M. Pirozhkova for the help with data processing and APRC computer group for the support.